\documentstyle[preprint,revtex]{aps}
\begin {document}
\draft
\preprint{UCI TR 92-17 /Uppsala U. PT11 1992}
\begin{title}
Proton $\beta$ Decay in Large Magnetic Fields
\end{title}
\author{Myron Bander\footnotemark\ }
\begin{instit}
Department of Physics, University of California, Irvine, California
92717, USA
\end{instit}
\author{H. R. Rubinstein\footnotemark\ }
\addtocounter{footnote}{-1}\footnotetext{e-mail:
mbander@ucivmsa.bitnet;
mbander@funth.ps.uci.edu}\addtocounter{footnote}{1}%
\footnotetext{e-mail:
rub@vand.physto.se}
\begin{instit}
Department of Radiation Sciences, University of Uppsala, Uppsala,
Sweden
\end{instit}
\receipt{April\ \ \ 1992}
\begin{abstract}
A delicate interplay between the anomalous magnetic moments of the
proton and neutron makes, in magnetic fields $B\ge 2\times 10^{14}$ T,
the neutron stable and for fields $B\ge 5\times
10^{14}$ T the proton becomes unstable to a decay into a neutron via
$\beta$ emission. Limits on the field strengths for which these
arguments hold are presented and are related to questions of vacuum
stability in the presence of such fields. Possible astrophysical
consequences are discussed.
\end{abstract}
\newpage
\narrowtext

\section{Introduction}

Very intense magnetic fields have been postulated to exist in
connection with some astrophysical objects. Fields with strengths
larger than $10^{14}$ T are associated with superconducting cosmic strings
\cite{Beres} and recently a proposal to explain
extragalactic gamma ray bursts involves fields of around $10^{13}$ T
\cite{Piran}.
We have been investigating several questions related to the
existence of such fields and to the behavior of elementary and
composite states in such environments. In a previous
paper \cite{Bander},  we discussed the possible breakdown
of constant magnetic fields with strengths beyond $10^{14}$ T.
Other mechanisms that destabilize strong
magnetic fields have been proposed \cite{proposals}, and here we add
another intriguing phenomenon to this list. For fields greater than
$5\times 10^{15}$ T the proton becomes heavier than the neutron and
decays into the latter by positron emission.

In Section II we study the behavior of an proton, neutron and electron
in an intense magnetic field and find the aforementioned amusing
result that the proton becomes unstable against neutron, positron and
neutrino decay. The decay rates and spectrum are obtained in Section
III. Whereas the discussion of the behavior of the electron
is on firm footing, questions may be raised as to the validity of our
treatment of the proton and neutron; these questions and the stability
of the vacuum in the presence of strongly interacting particles with
anomalous magnetic moments of non-electromagnetic origin are discussed
in Section IV. Conclusions and experimental consequences are presented
in the last section.

\section{Low Lying States for Particles in Uniform Magnetic Fields}
The quantum mechanics of a Dirac particle with no anomalous magnetic
moment in a uniform external magnetic field is straightforward. We shall
present the results for the case where particles do have such anomalous
moments. In reality, in fields so strong that the mass shifts
induced by such fields are of the order of the mass itself one
cannot define a magnetic moment as the energies are no longer linear in
the external field. Schwinger \cite{Schwinger} calculated the self
energy of an electron in an external field and we shall use
his results subsequently. We cannot follow this procedure for the proton
or  neutron as we do not have a good field
theory calculation of the magnetic moments of these particles, even for
small magnetic fields; all we have at hand is a phenomenological anomalous
magnetic moment. However, for fields that change the energies of these
particles by only a few percent, we will
consider these as point particle with the given anomalous moments. In
the Section IV we will discuss possible limitations of this approach.
\subsection{Protons in an External Field}  The Dirac Hamiltonian for a
proton with a uniform external magnetic field  ${\bf B} $ is
\begin{equation}
H=\mbox{\boldmath $\alpha$}\cdot\left ({\bf p}-e{\bf
A(r)}\right )+\beta M_p -
 {e\over {2M_p}}
\left ({g_p\over 2}-1\right )\beta{\bf\Sigma\cdot B}\, .\label{protonham}
\end{equation}
The vector potential ${\bf A(r)}$ is related to the field by ${\bf
A(r)}={1\over 2}{\bf r\times B}$ and $g_p=5.58$ is the proton's Land\'{e} g
factor. We first solve this equation for the case where the momentum along
the magnetic field direction is zero and then boost along that direction
till we obtain the desired momentum. For ${\bf B}$ along the ${\bf {z}}$
direction and $p_z=0$ the energy levels are \cite{Landaulevels}
\begin{equation}
E_{n,m,s}=\left [2eB(n+{1\over 2})-eBs+{M_p}^2 \right ]^{1\over 2}-
{e\over {2M_p}}\left ({g_p\over 2}-1\right )Bs\, .\label{Landauener}
\end{equation}
In the above, $n$ denotes the Landau level, $m$ the orbital angular
momentum about the magnetic field direction and $s=\pm 1$ indicates
whether the spin is along or opposed to that direction; the levels are
degenerate in $m$. $n=0$ and $s=+1$ yield the lowest energy
\begin{equation}
E={\tilde M}_p=M_p-{e\over {2M_p}}\left ({g_p\over 2}-1\right )B\, .
\label{protmass}
\end{equation}
As we shall be interested in these states only we will drop the $n$
and $s$ quantum numbers. The Dirac wave function for this
state is
\begin{equation}
\psi_{m,p_z=0}({\bf r})=\left ( \begin{array}{c} 1 \\0 \\0 \\0
\end{array}\right ) \phi_m({x,y})\, ,
\end{equation}
$\phi_m$'s are the standard wave functions of the lowest Landau
level;
\begin{equation}
\phi_m(x,y)={{\left [{1\over 2}|eB|\right ]^{{m+1}\over 2}}\over
{\sqrt{\pi m!}}}[x+iy]^m\exp\left [-{1\over 4}|eB|(x^2+y^2)\right ]\, .
\label{Landauwavefunction}
\end{equation}
Boosting to a finite value of $p_z$ is straightforward; we obtain
\begin{equation}
E_m(p_z)=\sqrt{{p_z}^2+{\tilde M}^2}\, ,
\end{equation}
with a wave function
\begin{equation}
\psi_{m,p_z}({\bf r})=\left ( \begin{array}{c} \cosh \theta \\0
\\ \sinh \theta \\0  \end{array}\right ) {{e^{ip_zz}}\over\sqrt{2\pi}}
\phi_m({x,y})\, ,\label{protonwavefunction}
\end{equation}
where $2\theta$, the rapidity, is obtained from $\tanh
2\theta=p_z/E_m(p_z)$.

In the non-relativistic limit the energy becomes
\begin{equation}
E_m(p_z)={\tilde M}+{{{p_z}^2}\over {2{\tilde M}}} \label{nrpe}
\end{equation}
and the wave function reduces to
\begin{equation}
\psi_{m,p_z}({\bf r})=\left ( \begin{array}{c} 1 \\0
\\ 0 \\0  \end{array}\right ) {{e^{ip_zz}}\over\sqrt{2\pi}}
\phi_m({x,y})\, . \label{nrpwf}
\end{equation}

\subsection{Neutrons in an External Field}
For a neutron the Dirac Hamiltonian is somewhat simpler
\begin{equation}
H=\mbox{\boldmath $\alpha$}\cdot{\bf p}+\beta M_n - {e\over {2M_n}}
\left ({g_n\over 2}\right )\beta{\bf\Sigma\cdot B}\, .\label{neutronham}
\end{equation}
with $g_n=-3.82$. Again for $p_z=0$ the states of lowest energy, the
ones we shall be interested in, have energies
\begin{equation}
E({\bf p_{\perp}}, p_z=0)={e\over {2M_n}}\left ({g_n\over 2}\right
)B+\sqrt{{\bf p_{\perp}}^2+{M_n}^2}
\end{equation}
Boosting to a finite $p_z$ we obtain
\begin{equation}
E({\bf p})=\sqrt{{E({\bf p_{\perp}}, p_z=0)}^2+{p_z}^2}\, .
\label{neutronener}
\end{equation}
The wave functions corresponding to this energy are
\begin{equation}
\psi_{\bf p}({\bf r})= {e^{i\mbox{\boldmath $p$}\cdot\mbox{\boldmath $r$}}
\over {(2\pi)^{3\over 2}}}u({\bf p},s=-1)\, ,
\end{equation}
where $u({\bf p},s=-1)$ is the standard spinor for a particle with
momentum ${\bf p}$, energy $\sqrt{{\bf p}^2+{M_n}^2}$ (not the energy
of Eq.~\ (\ref{neutronener})) and spin down.

In the non-relativistic limit
\begin{equation}
E({\bf p})=M_n+{e\over {2M_n}}\left ({g_n\over 2}\right )B+
{{{\bf p}^2}\over {2M_n}} \label{nrne}
\end{equation}
and the wave functions are
\begin{equation}
\psi_{\bf p}({\bf r})= \left ( \begin{array}{c} 0 \\1
\\ 0 \\0  \end{array}\right )
{e^{i\mbox{\boldmath $p$}\cdot\mbox{\boldmath $r$}}
\over {(2\pi)^{3\over 2}}}\, .\label{nrnwf}
\end{equation}

\subsection{Electrons in an External Field}
We might be tempted to use, for the electron, the formalism
used for the proton with the Land\'{e} factor replaced by $g_e=2+
\alpha/\pi$. However as we shall see for magnetic fields sufficiently
strong as to make the proton heavier than the neutron the change in
energy of the electron would appear to be larger than the mass of the
electron itself.
The point particle formalism breaks down and we have solve QED, to one
loop, in a strong magnetic field; fortunately this problem was treated
by Schwinger{\cite{Schwinger}}. The energy of an electron with $p_z=0$,
spin up and in the lowest Landau level is \begin{equation}
E_{m,p_z=0}=M_e\left [1+{\alpha\over {2\pi}} \ln \left ( {{2eB}\over
{{M_e}^2}} \right )\right ]\, .\label{electronener}
\end{equation}
For field strengths of subsequent interest this correction is negligible;
the energy of an electron in the lowest Landau level, with spin
down and a momentum of $p_z{\bf z}$ is
\begin{equation}
E_{m,p_z}=\sqrt{{p_z}^2+{M_e}^2}\, .
\end{equation}
and with wave function similar to those of the proton
\begin{equation}
\psi_{m,p_z}({\bf r})=\left ( \begin{array}{c} 0\\ \cosh \theta \\0
\\ \sinh \theta  \end{array}\right ) {{e^{ip_zz}}\over\sqrt{2\pi}}
{\phi_m^*}({x,y})\, ,
\end{equation}
where the boost rapidity, $2\theta$, is defined bellow
Eq. (\ref{protonwavefunction}) while the Landau level wave function is
defined in Eq. (\ref{Landauwavefunction}). The reason the complex
conjugate wave function appears is that the electron charge is
opposite to that of the proton.

\subsection{Decay Kinematics}
{}From Eq.~\ (\ref{protmass}) and Eq.~\ (\ref{nrne}) we note that the
neutron becomes stable against $\beta$-decay when the following inequality
is satisfied
\begin{equation}
-{e\over {2M_n}}\left ({g_n\over 2}\right )B
-{e\over {2M_p}}\left ({g_p\over 2}-1\right )B\ge M_n - M_p -
M_e\, ,
\end{equation}
or $B\ge 2\times 10^{14}$ T. On the other hand the proton becomes unstable for
decay into a neutron and a positron whenever
\begin{equation}
\-{e\over {2M_n}}\left ({g_n\over 2}\right )B
-{e\over {2M_p}}\left ({g_p\over 2}-1\right )B\sim 0.12\mu_NB  \ge M_n +
M_e - M_p\, ,\label{threshold}
\end{equation}
or for $B\ge 5\times 10^{14}$ T. We shall now turn to a calculation of
the life time of the proton in fields satisfying this inequality.
\section{Proton Life Time}
\subsection{Proton, Neutron and Electron Fields}
With the wave functions of the various particles in the
magnetic fields we may define field operators for these particles. For the
proton and electron we shall restrict the summation over states to the
lowest Landau levels with spin up, down respectively; for
magnetic fields of interest the other states will not contribute to the
calculation of decay properties. For the same reason, the neutron
field will be restricted to spin down only. The proton and neutron
kinematics will be taken as non-relativistic.
\begin{eqnarray}
\Psi_p({\bf r})=\sum_m\int dp_z\left [a_m(p_z)\left ( \begin{array}{c} 1 \\0
\\ 0 \\0  \end{array}\right ) {{e^{ip_zz}}\over\sqrt{2\pi}}
\phi_m({x,y})\right. \nonumber\\
 \left.
+{b^{\dag}}_m(p_z)\left ( \begin{array}{c} 0 \\0
\\ 1 \\0  \end{array}\right ){{e^{-ip_zz}}\over\sqrt{2\pi}}
\phi_m({x,y})\right ]\, ,
\label{protonfield}
\end{eqnarray}
with $\phi_m({x,y})$ defined in Eq.~\ (\ref{Landauwavefunction})
and the energy,
$E_m(p_z)$ in Eq.~\ (\ref{nrpe}). $a_m(p_z)$ is the annihilation operator
for a proton with momentum $p_z{\bf z}$ and angular momentum $m$;
$b_m(p_z)$ is the same for the negative energy states. For the
neutron the field is
\begin{equation} \Psi_n({\bf r})=\int d^3p \left
[a({\bf p})\left ( \begin{array}{c} 0 \\1  \\ 0 \\0  \end{array}\right )
{e^{i\mbox{\boldmath $p$}\cdot\mbox{\boldmath $r$}} \over {(2\pi)^{3\over
2}}} +b^{\dag}({\bf p})\left ( \begin{array}{c} 0 \\0  \\ 0 \\1
\end{array}\right )  {e^{-i\mbox{\boldmath $p$}\cdot\mbox{\boldmath $r$}}
\over {(2\pi)^{3\over 2}}}\right ]\, ,\label{neutron field}
\end{equation}
with an obvious definition of the annihilation operators.
For the electron we use fully relativistic kinematics and the field is
\begin{eqnarray}
\Psi_e({\bf r})=\sum_m\int dp_z \sqrt{{M_e}\over E}\left [a_m(p_z)
\left ( \begin{array}{c} 0\\ \cosh \theta \\0
\\ \sinh \theta  \end{array}\right ) {{e^{ip_zz}}\over\sqrt{2\pi}}
{\phi_m^*}({x,y})\right. \nonumber\\ \left.
+b^{\dag}_m(p_z)
\left ( \begin{array}{c} 0\\ \cosh \theta \\0
\\ \sinh \theta  \end{array}\right ) {{e^{-ip_zz}}\over\sqrt{2\pi}}
{\phi_m^*}({x,y})\right ]\, .\label{electron field}
\end{eqnarray}
\subsection{Decay Rates and Spectrum}
The part of the weak Hamiltonian responsible for the decay
$p\rightarrow n+e^{+}+\nu_e$ is
\begin{equation}
H={{G_F}\over {\sqrt{2}}}\int d^3x \bar{\Psi}_n\gamma_{\mu}(1+\gamma_5)\Psi_p
\bar{\Psi}_{\nu}\gamma^{\mu}(1+\gamma_5)\Psi_e\, .
\end{equation}
For non-relativistic heavy particles the matrix element of this Hamiltonian
between a proton with quantum numbers $p_z=0$ , $m=m_i$, a neutron
with momentum
${\bf p}_n$, a neutrino with momentum  ${\bf p}_{\nu}$ and an electron in state
$m=m_f$ and with ${p_{z,e}}$ is
\begin{eqnarray}
\langle H\rangle ={{2G_F}\over {(2\pi)^3}}\left( {{E_e+{p_{z,e}}}\over
{E_e-{p_{z,e}}}}\right )^{1\over 4}\sin (\theta_{\nu}/2)
\sqrt{M_e\over {E_e}}\delta (p_{z,e}+p_{z,\nu}+p_{z,n})\nonumber\\
\int dx\, dy\, {\phi_{m_f}^*}({x,y}){\phi_{m_i}}({x,y})\exp
[-i(\mbox{\boldmath  $p_{\perp,n}+p_{\perp,\nu}$})\cdot\mbox{\boldmath
$r_{\perp}$}]\, ;\label{matelem}
\end{eqnarray}
$\theta_{\nu}$ is the azimuthal angle of the neutrino. The integral in
the above
expression can be evaluated in a multipole expansion. Note that the natural
extent of the integral in the transverse direction is $1/\sqrt{eB}$ whereas the
neutron momenta are, from Eq.~\ (\ref{threshold}), of the order of
$\sqrt{0.12eB}$; thus setting the exponential term in this integral
equal to one
will yield a good estimate for the rate and spectrum of this decay.
The positron spectrum is given by
\begin{equation}
{{d\Gamma}\over {dp_{z,e}}}={4\over 3}{{G_F^2 M_p}\over {(2\pi)^6}}
{{E_e+p_{z,e}}\over {E_e}}(\Delta-E_e)^3\, ;
\end{equation}
where $\Delta= 0.12\mu_NB-M_n+M_p$. For $\Delta\gg M_e$ the total rate
is easily obtained
\begin{equation}
\Gamma={2\over 3}{{G_F^2 M_p}\over {(2\pi)^6}}\Delta^4\, .
\end{equation}
The lifetime is $\tau\sim 1.5\times 10^2 (10^{15}{\rm T}/B)^4$ s.

\section{Field Stability in the Presence of Composite Hadrons}
In Refs. \cite{Bander,proposals} it was noted that vector particles
with anomalous magnetic moments induce instabilities for large magnetic
fields. Such a mechanism is not induced by spin one-half fields, even
with anomalous moments. It is however clear that the discussion of the
previous section regarding protons and neutrons has to break down at a
sufficiently strong field. Let us look specifically at the neutron
case. From Eq.~(\ref{neutronener}) we see that for $M_n-\mu B\le 0$
positive and negative energy levels cross and pair creation becomes
possible. Following the analysis of Ref. \cite{Bander} we
can show that the vacuum does not decay into neutron-antineutron
pair. (We cannot rule out the possibility that the strong interactions
responsible for the anomalous moment could induce, for example, pion
pair creation.) What occurs, is that the vacuum becomes a linear
combination of the old vacuum and two fermion-antifermion pairs; a
mixing with a single fermion-antifermion pair is ruled out on parity
grounds.

It is amusing to look at this problem in two dimensions\footnotemark
\footnotetext{This
two dimensional
argument is due to A. Schwimmer.}. A Lagrangian
for a charged Fermion with an anomalous moment is
\begin{equation}
{\cal L}=i\overline{\psi}\gamma_\mu\left (\partial^\mu-eA^\mu\right
)\psi +m\overline{\psi}\psi+
\mu\overline{\psi}\sigma_{\mu\nu}\psi F^{\mu\nu}\, .
\end{equation}
The bosonized version of this Lagrangian is \cite{boson}
\begin{equation}
{\cal L}={1\over 2}\partial_{\mu}\phi\partial^{\mu}\phi +
m\left [1-\cos (2\sqrt{\pi}\phi)\right ]
+eE\phi+\frac{1}{2} \mu E\sin (2\sqrt{\pi}\phi)\, .
\end{equation}
In two dimensions there is only one field component, $F^{0,1}=E$. We
note that for non-zero charge, $e$, a true instability develops,
whereas for $e=0$ and non-zero $\mu$ the field $\phi$ acquires a
finite expectation value, corresponding to fermion pairs in the
original model.

{}From the above we expect the arguments of the previous sections to
break down for critical fields, $B_c\sim M_n^2/e\sim 10^{16}$ T.
However, the composite nature of the nucleons will cause a breakdown
for lower fields. A reasonable estimate would be a critical field for
the constituent quarks, $B\sim M_q^2/e\sim 10^{15}$ T. Thus our
analysis should be valid for fields below this value and that leaves a
window, $5\times 10^{14}\ {\rm T}\le B\le 10^{15}\ {\rm T}$ where we
expect the proton to be heavier than the neutron and to decay into it.

\section{Conclusions and Experimental Consequences}
We studied the mass evolution of protons, neutrons and electrons
in strong magnetic field and concluded that the proton
will decay $\beta$ decay, $p\rightarrow n+e^{+}+\nu_e$, in a
sufficiently strong external magnetic field.
The new mechanism for positron generation might have astrophysical interest.
There is indeed an overabundance of
positrons as compared with that accounted for by existing
mechanisms \cite{Cesarsky}.
\section{Acknowledgments}

We thank Miriam Leurer and Adam Schwimmer as well as other members of the
Weizmann Institute for interesting discussions.

\end{document}